\documentclass[iop]{emulateapj}
\usepackage{color}
\usepackage{amsmath}

\newcommand{\Teff}{\ensuremath{T_\mathrm{eff}}}
\newcommand{\vsini}{\ensuremath{v \sin i}}

\begin{document}
\title{K2-136: A Binary System in the Hyades Cluster Hosting a Neptune-Sized Planet}

\shorttitle{K2-136: A Hyades Binary Star with a Neptune-sized Planet}
\shortauthors{Ciardi et al.}

\author{David R.~Ciardi\altaffilmark{1},
Ian J.~M.~Crossfield\altaffilmark{2},
Adina D.~Feinstein\altaffilmark{3}, 
Joshua E.~Schlieder\altaffilmark{4}, 
Erik A.~Petigura\altaffilmark{5},
Trevor J.~David\altaffilmark{6},
Makennah Bristow\altaffilmark{7},
Rahul I.~Patel\altaffilmark{1},
Lauren Arnold\altaffilmark{8}, 
Bj\"orn Benneke\altaffilmark{9},
Jessie L.~Christiansen\altaffilmark{1},
Courtney D.~Dressing\altaffilmark{10},
Benjamin J.~Fulton\altaffilmark{6},
Andrew W.~Howard\altaffilmark{6},
Howard Isaacson\altaffilmark{10}
Evan Sinukoff\altaffilmark{6,11},
Beverly Thackeray\altaffilmark{12,13}\\
}                             
\altaffiltext{1}{Caltech/IPAC-NASA Exoplanet Science Institute Pasadena,  CA USA}
\altaffiltext{2}{Department of Physics, Massachusetts Institute of Technology, Cambridge, MA USA}
\altaffiltext{3}{Department of Physics and Astronomy, Tufts University, Medford, MA USA}
\altaffiltext{4}{Exoplanets and Stellar Astrophysics Laboratory, Code 667, NASA Goddard Space Flight Center, Greenbelt, MD USA}
\altaffiltext{5}{Division of Geological and Planetary Sciences, California Institute of Technology, Pasadena, CA USA}
\altaffiltext{6}{Cahill Center for Astrophysics, California Institute of Technology, Pasadena, CA USA}
\altaffiltext{7}{Department of Physics, University of North Carolina at Asheville, Asheville, NC USA}
\altaffiltext{8}{Center for Marine and Environmental Studies, University of the Virgin Islands, Saint Thomas, United States Virgin Islands, USA}
\altaffiltext{9}{Universit\'e of Montr\'eal, Montr\'eal, Qu\'ebec, CA}
\altaffiltext{10}{University of California at Berkeley, Berkeley, CA 94720, USA}
\altaffiltext{11}{Institute for Astronomy, University of Hawai`i at M\={a}noa, Honolulu, HI USA}
\altaffiltext{12}{Department of Physics, California State University San Bernardino, San Bernardino, CA USA}
\altaffiltext{13}{Department of Astronomy, University of Maryland College Park, College Park, MD USA}

\email{ciardi@ipac.caltech.edu}

\slugcomment{Accepted for publication to AAS Journals}

\begin{abstract} 
We report the discovery of a Neptune-size planet ($R_p = 3.0 R_\oplus$) in the Hyades Cluster.  The host star is in a binary system, comprising a K5V star and M7/8V star with a projected separation of 40 AU.  The planet orbits the primary star with an orbital period of 17.3 days and a transit duration of 3 hours. The host star is bright ($V=11.2$, $J=9.1$) and so may be a good target for precise radial velocity measurements.   K2-136A~c is the first Neptune-sized planet to be found orbiting in a binary system within an open cluster. The Hyades is the nearest star cluster to the Sun, has an age of 625-750 Myr, and forms one of the fundamental rungs in the distance ladder; understanding the planet population in such a well-studied cluster can help us understand and set constraints on the formation and evolution of planetary systems.
\end{abstract}


\section{Introduction}\label{sec:intro}
Most stars are thought to form in open clusters \citep{ll2003}, but most planets have been found around old, isolated stars that have long since left their nascent cluster families.  There have a been a series of studies to try to find planets in open clusters.  Part of the reason to study planets in open clusters is that the stars are typically well understood in terms of mass, metallicity, and age (especially in comparison to field stars), and, therefore,  because the derivation of planet parameters requires accurate and precise knowledge of the host stars, any planets found within open clusters would also be much better understood. With the pending release of Gaia distances, most field star planetary systems will be more clearly defined akin to what is currently possible with systems in open clusters -- with the exception of ages.  The discovery of exoplanets in open clusters enable us to explore possible evolutionary effects on the distribution and characteristics on exoplanets as a function of time and age.

While there are more than 1300 confirmed exoplanets with mass determinations and more than 2200 statistically validated planets \citep{NEA,akeson2013}, only about $\sim$1\% have been discovered in open clusters, and the majority of these are Jupiter-sized planets.  The first planet discovered in any open cluster was in the Hyades; $\epsilon$ Tauri b is a $\approx 7M_{Jup}$ mass planet in a 600 day orbit around an evolved K0 giant star \citep{sato2007}.  Since that discovery, there have been a handful of planets discovered via radial velocity in young open clusters including an additional planet in the Hyades [\cite[HD285507 b,][]{quinn2014}], two planets in the Taurus region [\cite[V830 Tau b,][]{donati2016}; \cite[Cl Tau b,][]{jk2016}], one planet in the distance cluster NGC2423 [\cite[NGC2423-3 b,][]{lm2007}], and three planets in Praesepe Cluster [\cite[Pr0201 b, Pr0211 b,][]{quinn2012}; \cite[Pr0211 c,][]{malavolta2016}]. However, most transit cluster surveys prior to the Kepler mission were not sensitive enough or had samples large enough to detect the more common Neptune-sized and smaller planets \citep[e.g.,][]{pg2006, quinn2012, quinn2014, brucalassi2017}. 

With Kepler and K2, a handful of transiting small planets have been discovered in open clusters.  Kepler was sensitive enough to detect two super-Earth-sized planets in the billion-year-old cluster NGC 6811, located more than 1000~pc away \citep[Kepler-66b, $2.80$R$_\oplus$; Kepler-67b, $2.94$R$_\oplus$;][]{meibom2013}. K2, through its larger survey area, has been surveying open clusters much closer to home \citep{howell2014}. With K2, a sub-Saturn-sized planet (K2-33b, $R=5.04$R$_\oplus$) was discovered in the  5-10 Myr old cluster Upper Scorpius \citep{david2016b,mann2016}; six planets spanning super-Earth to Neptune-sized (K2-95b, K2-100b, 101b, 102b, 103b, 104b) have been detected orbiting K and M dwarfs and in the Praesepe Cluster \citep{obermeier2016,mann2017}, and a Neptune-sized planet (K2-25b, $R=3.47$R$_\oplus$) was discovered orbiting an M4.5V star in the Hyades \citep{mann2016,david2016a}. 

A key goal of young cluster exoplanet searches is to test whether planets around young cluster stars have the same occurrence distribution as mature planets around field stars \citep[e.g.,][]{meibom2013}; this would be a relatively expected result, if field stars are indeed primarily born in clusters.  Thus, understanding the frequency and distribution of planets in open clusters -- particularly those that are younger than the field stars -- can help constrain the formation and evolution mechanisms that shape the frequency and distribution of planets observed in the older field stars.

The Hyades is the nearest open cluster to the Sun and is one of the best studied clusters.   The cluster center is located $46.34\pm0.27$ pc away \citep{vanleeuwen2009}, but the cluster members themselves span an extent that is 10-20~pc across \citep[e.g.,][]{mann2016}. The Hyades has a metallicity slightly higher than solar ([Fe/H] $\approx 0.13\pm0.01$; \citealt{paulson2003,maderak2013}), and typically, the age of the Hyades is quoted as $625\pm50$ Myr \citep{perryman1998}, although some recent work indicates that the cluster may be slightly older \citep[$750\pm100$ Myr;][]{bh2015,david2015}. The stellar binarity rate within the Hyades is also fairly well documented showing a strong dependence on stellar type; stars earlier than solar have nearly a 100\% companion fraction and that fraction drops to below 50\% for early-K stars \citep{bv2007}.

\begin{figure*}[ht!]
\vspace{-0.1in}
\begin{center}
\includegraphics[scale=0.75]{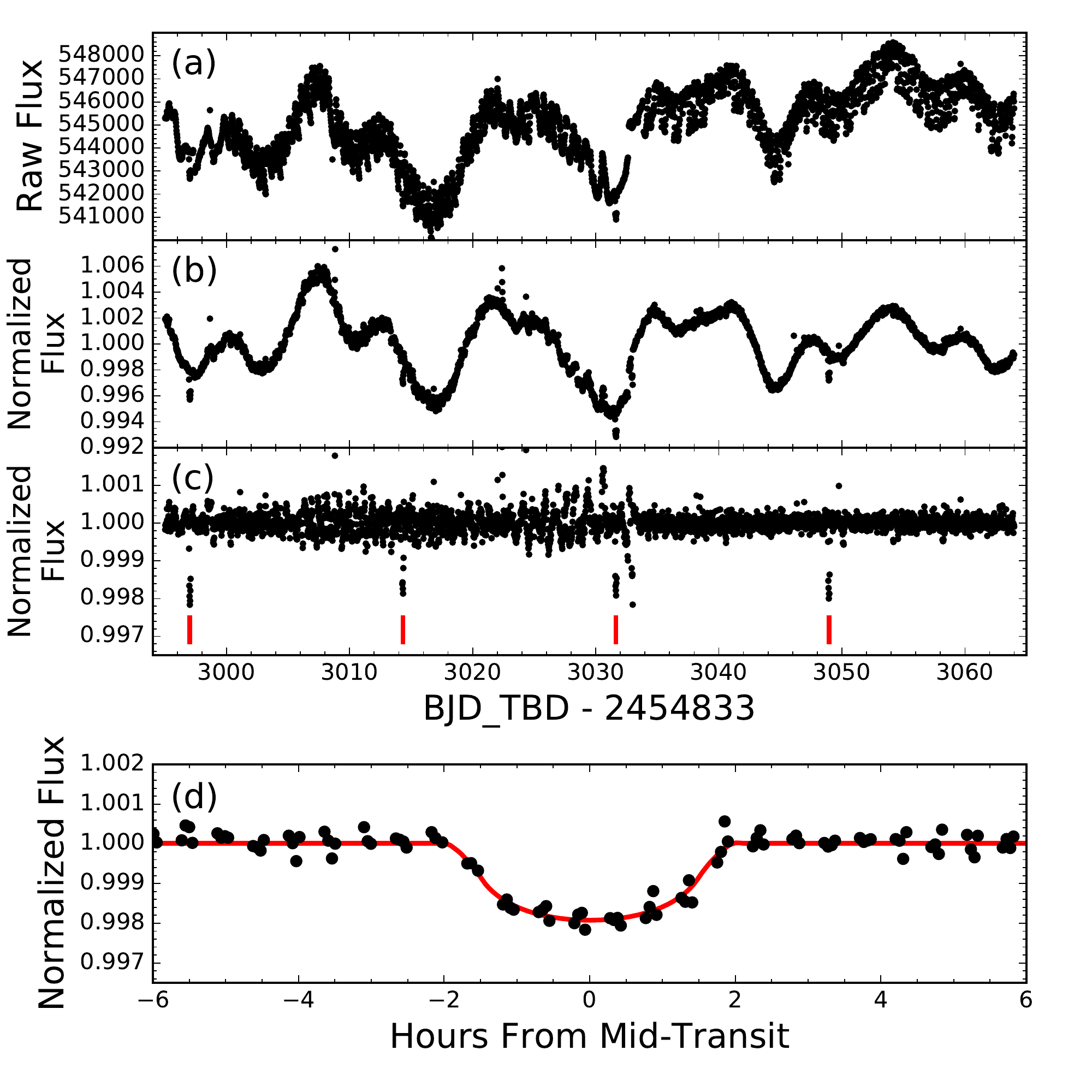}
\caption{\label{fig:fits} Our K2 photometry: (a) immediately after extraction from the pixel-level data; (b)  after removal of systematics, showing the stellar variability; (c) after smoothing and detrending, with vertical ticks indicating the locations of  transits; and (d)  the phase-folded photometry with the best-fit transit model fit to the light curve.  The feature at time index 3033 is a residual systematic induced by K2's motion. }
\end{center}
\end{figure*}

Recent work has suggested that the presence of stellar companions may inhibit the formation of planets \citep[e.g.,][]{kraus2016}, but other work has suggested the stellar companion rate of planet hosting stars is similar to the field star companion rate \citep[e.g.,][]{horch2014}.  Additionally, stellar encounters (i.e., fly-bys and collisions) within the cluster environment may alter the formation and/or survival of planets and planetary systems, in comparison to what might be expected for single isolated stars \citep[e.g.,][]{malmberg2011}.  Finding planets within the Hyades cluster can help yield important constraints on planet formation and evolutionary theories -- particularly, if the frequency of planets in the Hyades as a function of stellar type and stellar multiplicity can be established.

This paper presents the discovery of a Neptune-sized planet host by the K-dwarf EPIC~247589423 within the Hyades cluster. The detection was made with K2; we have performed a suite of follow-up observations which include high-resolution imaging and spectroscopy. In addition to the transit light curve of the planet, the imaging was used to detect a late M-dwarf stellar companion; spectroscopy was utilized to derive precise stellar parameters of the primary host star and show that the star is indeed a Hyades member, based upon kinematic arguments. The primary star is a K5V and has an M7/8V stellar companion located approximately 40 AU (projected) from the primary star.  The light curve modeling and validation is consistent with a Neptune-sized planet ($\sim3.0$ R$_\oplus$) orbiting the primary star with a period of $\sim17.3$ days. We demonstrate that the primary K5V star is the host of the planet. With the discovery of the stellar companion and the planet, we set the nomenclature of the system: K2-136A is the K5V primary star; K-136B is the M7/8V stellar companion. Finally, we should that the Neptune-sized planet most likely orbits the primary star.

\section{K2 Detection}\label{sec:k2detect}
EPIC~247589423 (LP~358-348) was observed by K2  at a 30-minute cadence in Campaign 13, which ran from 2017 March 08 until 2017 May 27.  The star was proposed for observation by numerous K2 General Observer programs:  13008, A.\ Mann; 13018, I.\ Crossfield; 13023, L.\ Rebull; 13049, E.\ Quintana; 13064, M.\ Agueros; 13077, M.\ Endl; and 13090, J.\ Glaser.  The properties of EPIC~247589423 are summarized in Table~\ref{tab:stellar}. 

We identified the transit candidate in the light curve analysis of raw K2 cadence data using a series of free software tools made available by the community, following the same approach described by \citet{crossfield2016} and Christiansen et al.\ (in review). In brief: we processed the cadence data into target pixel files with \texttt{Kadenza}\footnote{\url{https://github.com/KeplerGO/kadenza}} \citep{kadenza}, generated time-series photometry and removed K2's well-known systematics using \texttt{k2phot}\footnote{\url{https://github.com/petigura/k2phot}}, and searched for  candidate planet transits using \texttt{TERRA}\footnote{\url{https://github.com/petigura/terra}}
\citep{petigura2013b,petigura2013a}.  Fig.~\ref{fig:fits} shows the several stages of light curve processing. 

The resulting light curve shows coherent variation with a peak-to-peak amplitude of roughly 1\%, as seen in Fig.~\ref{fig:fits}c.  \texttt{TERRA} also identified one strong transit-like signal clearly visible in Fig.~\ref{fig:fits} with $P\approx17.3$~d, a depth of $\sim 1500$~ppm, and with a S/N=18. We saw no obvious secondary eclipse ($\lesssim 240$ ppm) or evidence of flux modulation on the detected period. After masking out those transits, \texttt{TERRA} found no other transit signals with S/N\,$\ge$\,7.  

\section{Follow-Up Observations}\label{sec:fop}
Following the detection of the candidate planet around EPIC 247589423 in the K2 light curve, we began our standard follow-up process to assess the stellar parameters of the targets and to validate the planetary candidate as a true planetary system utilizing both archival data and new imaging and spectroscopy data \citep[e.g.,][]{crossfield2016, martinez2017, dressing2017a, petigura2017}.

\begin{figure}[!ht]
\begin{center}
\includegraphics[angle=0,scale=0.65,keepaspectratio=true]{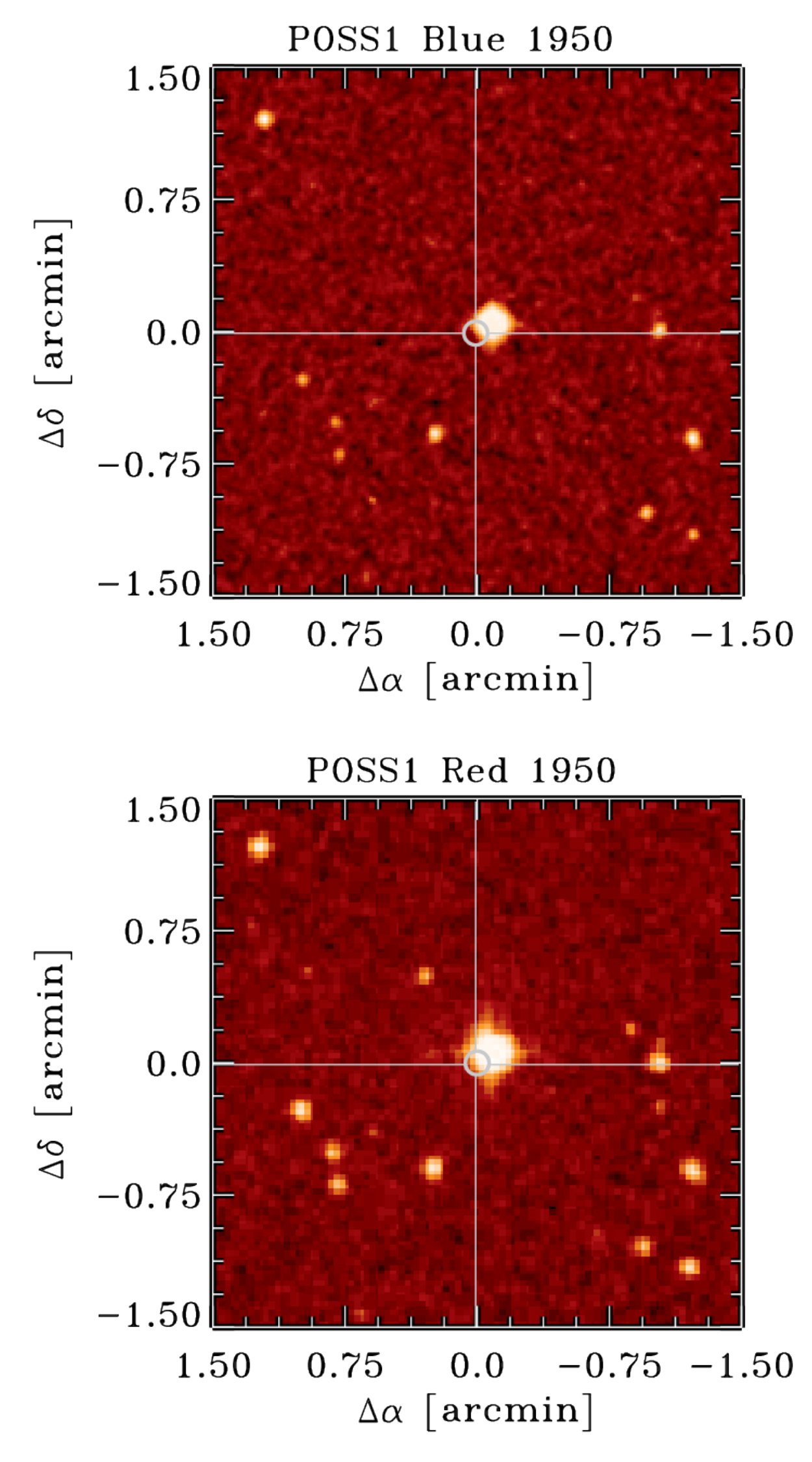}
\caption{POSS1 Blue and Red plates observed in 1950. The circle shows the location of EPIC~247589423 at the 2017 position of the star. Between 1950 and 2017, the star moved moved by $\sim 6\arcsec$, which can be clearly seen in the POSS images. The POSS1 plate rules out a background star coincident with the current location of EPIC~247589423 to $\Delta B\sim 3$ mag and $\Delta R\sim 4$ mag. \label{fig:pm}}
\end{center}
\end{figure}
\subsection{Archival Imaging and Proper Motion}\label{subsec:pm}
EPIC~247589423 is a high proper motion star \citep[$+81.8$ mas yr$^{-1}$ in right ascension and $-35.2$ mas yr$^{-1}$ in declination; UCAC4, ][]{zacharias2013}. In the 67 years since the 1950 Palomar Observatory Sky Survey (POSS) images, EPIC~247589423 has moved more than 6$\arcsec$, enabling us to utilize archival POSS data to search for background stars that are now, in 2017, hidden by EPIC~247589423.  The Blue POSS1 image has better resolution ($\sim 2\arcsec$ \textit{vs.} $\sim 4\arcsec$) but the Red POSS1 image goes deeper ($\Delta \mathrm{mag} \sim 4$ \textit{vs.} $\Delta \mathrm{mag} \sim 6$)
 
Using the 1950 POSS data (Fig.~\ref{fig:pm}), we find no evidence of a background star at the current position of EPIC~247589423 to a differential magnitude of $\Delta B\sim 3$ mag in the blue and $\Delta R\sim 4$ mag in the red. Because EPIC~247589423 is slightly saturated in the POSS images, this sensitivity was estimated by placing fake sources at the epoch 2017 position of EPIC~247589423 in the epoch 1950 images and estimating the 5$\sigma$ threshold for detection. The photometric scale of the image (and hence, the magnitudes of the injected test stars) was set using the star located 40\arcsec\ to the southeast of EPIC~247589423, which has an optical magnitude of approximately $B\approx 17$ mag and $R\approx16$ mag. 

This analysis rules out a $\lesssim10\%$ eclipsing binary that is $3-4$ magnitudes fainter than the primary star, but it does not rule out the more extreme background eclipsing binaries (a 100\% eclipsing binary could produce a 1500~ppm transit at a differential magnitude of $\sim 7$ magnitudes).  However, this analysis was sufficient for us to initiate the remainder of the follow-up observations.

\begin{figure*}[!ht]
\begin{center}
\includegraphics[angle=0,scale=0.75,keepaspectratio=true]{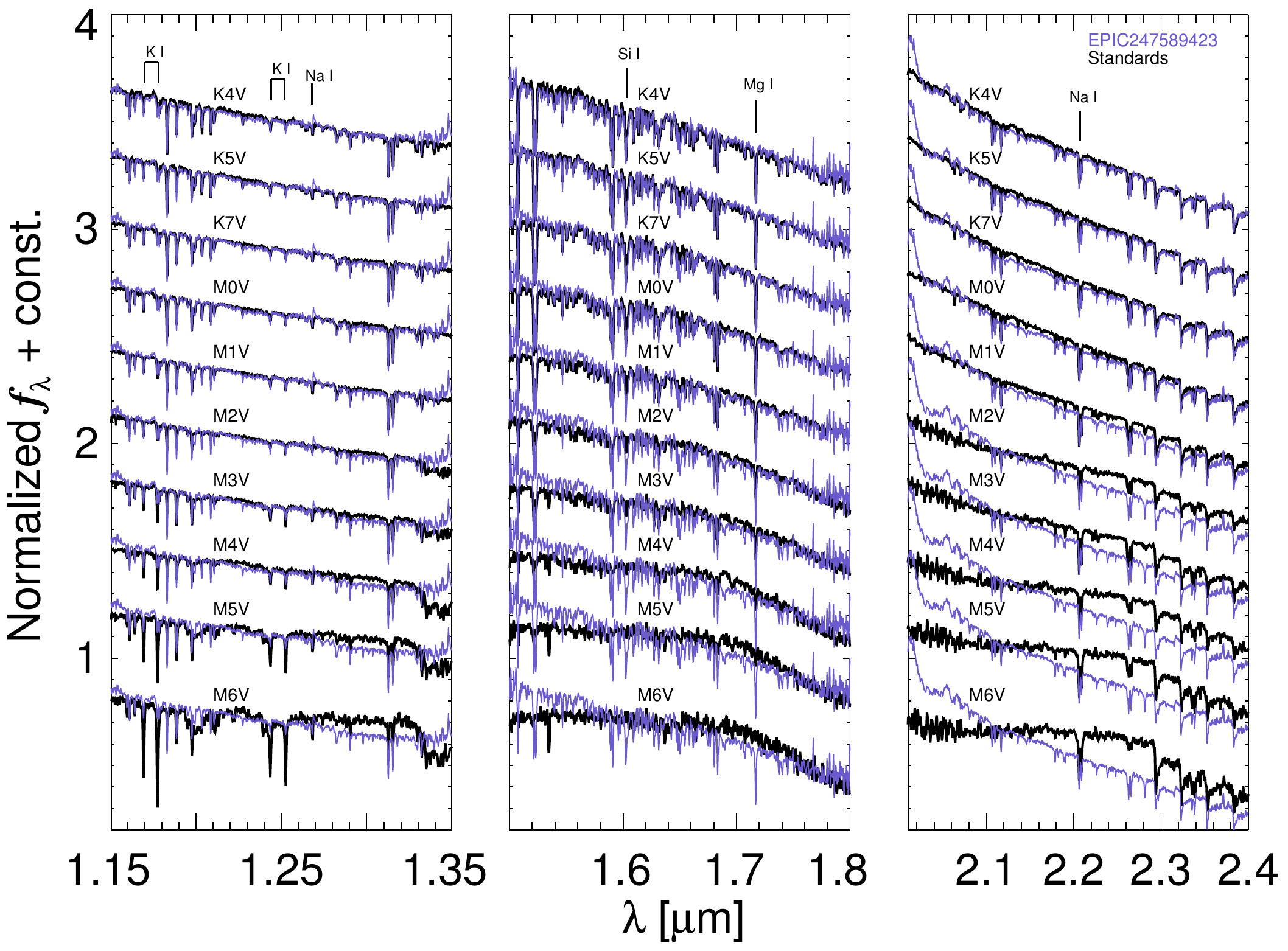}
\caption{JHK-band IRTF/SpeX spectra of K2-136 (EPIC 247589423) compared with late-type dwarf standards from the IRTF spectral library. All spectra are normalized to the continuum in each of the plotted regions. The star is a best visual match to spectral type $\sim$K5 across the three near-IR bands. \label{fig:spex}}
\end{center}
\end{figure*}

\subsection{Spectroscopy}\label{subsec:spec}
We performed both near-infrared and optical spectroscopy in order to characterize the host star properties and to search for secondary spectral lines. 

\subsubsection{IRTF SpeX}\label{subsubsec:irtf}
We observed EPIC~247589423 with the near-infrared cross-dispersed spectrograph SpeX \citep[][]{rayner2003,raynor2004} on the 3m NASA Infrared Telescope Facility on 2017 July 24 UT (Program 2017A019, PI C.\ Dressing). While available photometry indicates that the star is late-type, follow-up spectroscopy is essential to measure the spectral type and fundamental parameters. 

We observed EPIC~247589423 under clear skies with an average seeing of $\sim$ 0.7$^{\prime\prime}$. We used SpeX in its short cross-dispersed mode (SXD) with the 0.3$\times$15$^{\prime\prime}$ slit, allowing us to observe the star over $0.7-2.55\ \mu$m  at resolution $R \sim 2000$. The target was observed at two locations along the slit in three AB nod pairs using a 50s integration time in each frame, providing a total integration time of 300s. The slit position angle was synced to the parallactic angle to avoid differential slit losses.  An A0 standard, HD31411, was observed after our target and flat and arc lamp exposures were taken immediately after that, to allow for telluric correction and wavelength calibration using the data reduction package, SpeXTool \citep{vacca2003,cushing2004}.

SpeXTool performs flat fielding, bad pixel removal, wavelength calibration, sky subtraction, flux calibration, and spectral extraction and combination. The final extracted and combined spectra have signal-to-noise ratios (SNR) of 175 per resolution element in the $J$-band (1.25$\mu$m), 217 per resolution element in the $H$-band (1.6$\mu$m), and 208 per resolution element in the $K$-band (2.2$\mu$m). The $JHK$-band spectra were compared to late-type standards from the IRTF Spectral Library \citep{rayner2009}, seen in Figure~\ref{fig:spex}. EPIC~247589423 is an approximate visual match to the K5 standard across all three bands. The increased noise visible in the regions of strong H$_2$O absorption is a result of increased telluric contamination, potentially due to the relatively large $\sim$19$^{\circ}$ separation between the primary target and the available A0 standard. 

Following the methods presented in \cite{mann2013}, we use our SpeX spectrum  to estimate the fundamental parameters of effective temperature (\Teff), radius ($R_*$), mass ($M_*$), and luminosity ($L_*$) for EPIC~247589423. We used the index-based temperature relations of \citet{mann2013} to estimate the temperature in each of the $J$-, $H$- and $K$-bands and calculated the mean of the three values. We estimated the uncertainty by adding in quadrature the standard deviation of the mean and the scatter in each of the \citet{mann2013} index relations. The resulting \Teff\ = 4360 $\pm$ 206~K was then used to estimate the remaining stellar parameters and their uncertainties using the polynomial relations from \citet{mann2013}. We estimate $R_*/R_\odot$ = 0.674 $\pm$ 0.061, $M_*/M_\odot$ = 0.696 $\pm$ 0.070, and $L_*/L_\odot$ = 0.152 $\pm$ 0.052. The radius and mass yield an empirical stellar density of $3.2 \pm 1.0$~g~cm$^{-3}$. 

We also used the TiO5 and CaH3 molecular indices from \citet{lepine2003} to measure a spectral type of K7 $\pm$ 0.5. This visible index based spectral type is consistent with the estimated stellar parameters \citep{pecaut2013} and the visual comparison to standards in the near-IR. Given these results, we adopt a conservative dwarf spectral type of K5 $\pm$ 1 for EPIC~247589423. The late-M type companion detected at close separation in near-IR adaptive optics imaging is $\sim$5-10 magnitudes fainter than the K5 star at visible to near-IR wavelengths (see \S \ref{subsec:hri}). Thus, it only contributes  $\lesssim$1\% of the flux across the wavelength ranges used in the SpeX analyses and does not significantly affect the results.

\subsubsection{Keck HIRES}\label{subsubsec:hires}
We also observed the star on UT 2017 Aug 04 with the HIRES spectrometer \citep{vogt1994} on the Keck~I telescope. We observed for 64~s using the C2 decker and no iodine cell, achieving 10,000 counts on the HIRES exposure meter (corresponding to S/N of 22~pix$^{-1}$ on blaze). As an independent check of the SpeX derived values, stellar parameters were estimated from the iodine-free template spectrum using the SpecMatch-Emp code \citep{yee2017}\footnote{https://github.com/samuelyeewl/specmatch-emp}.  

SpecMatch-Emp contains a dense spectral library of $\sim$400 touchstone stars with well-determined properties. This library is made up of HIRES spectra taken at high signal to noise ($SNR > 100$/pix).  SpecMatch-Emp fits an unknown target spectrum by finding the optimum linear combination of library spectra that best matches the target spectrum. SpecMatch-Emp performs particularly well when analyzing cool stars with $\Teff < 4700$~K (SpT $\ge$ K4). At low temperatures, the onset of dense molecular bands challenges LTE spectral synthesis codes. SpecMatch-Emp achieves an accuracy of 70 K in \Teff, 10\% in $R_∗$, and 0.12 dex in [Fe/H] \citep{yee2017}. Because the spectral library radii are measured using model-independent techniques such as interferometry or spectrophotometry, the derived radii do not suffer from model-dependent offsets associated with converting \Teff, $\log g$, and [Fe/H] into $R_*$. 

The HIRES SpecMatch-Emp results are consistent with the SpeX results and the adopted spectral type of K5 $\pm$ 1; we find  $\Teff=4364 \pm 70$~K, $R_*=0.71\pm0.10\ R_\odot$, [Fe/H]\,=\,$+0.15\pm0.09$. We note that the stellar metallicity is consistent with the Hyades cluster metallicity of [Fe/H]$ = 0.13$. We use the \texttt{isochrones} package \citep{morton2015} to convert the SM-Emp stellar parameters (\Teff, $R_*$, and [Fe/H]) and the $K_s$ magnitude into a stellar mass and $\log g$.  With these inputs, we find $M_*=0.71 \pm 0.06 M_\odot$ and $\log g=4.63 \pm 0.11$. We also used the HIRES spectrum to measure the star's radial velocity, RV = 39.6 $\pm$ 0.2 km s$^{-1}$, and projected rotational velocity, $v\mathrm{sin}i = 3.9\pm1.0$ km s$^{-1}$ (see Table~\ref{tab:stellar}. 

To search for  stellar companions at small separations, we ran the secondary line search algorithm presented by \citet{kolbl2015} on the HIRES spectrum. There is no evidence of secondary lines in the spectrum for companions down to $\Delta V\lesssim5$~mag and $\Delta$RV$\gtrsim$10 km s$^{-1}$.  These results complement the results of the high resolution imaging where the spectroscopy can probe regions inside the inner working angle of the imaging.  The results are also consistent with the results of the infrared high-resolution imaging presented in the next section where a late M-dwarf has been detected. That M-dwarf would be $\sim$10 magnitudes fainter than the K5V star in the $V$-band and beyond the sensitivity of the HIRES spectrum.

\begin{figure}[!ht]
\begin{center}
\includegraphics[angle=0,scale=0.275,keepaspectratio=true]{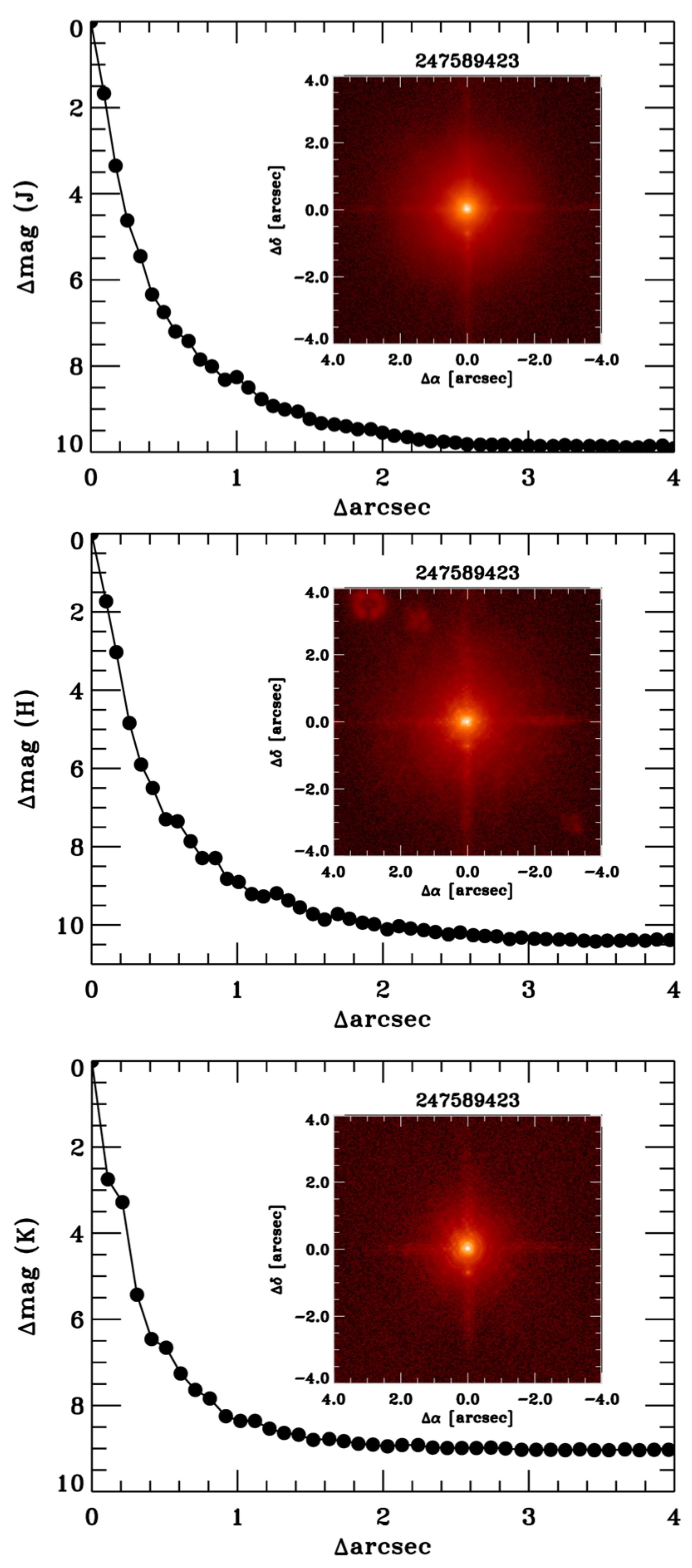}
\caption{Contrast sensitivities and inset images of EPIC~247589423 in the J, H, and $Br$-$\gamma$ filters as observed with the Palomar Observatory Hale Telescope adaptive optics system; the secondary companion $\sim 0.72\arcsec$ to the south of the primary target is clearly detected.  The $5\sigma$ contrast limits for additional companions, in $\Delta$magnitude, are plotted against angular separation in arcseconds for each of the filters.  The black points represent one step in the FWHM resolution of the images.  \label{fig:ao_contrast}}
\end{center}
\end{figure}

\subsection{High-resolution Imaging}\label{subsec:hri}
As part of our standard process for validating transiting exoplanets, we observed EPIC~247589423 with infrared high-resolution adaptive optics (AO) imaging, both at Keck Observatory and Palomar Observatory.  The Keck Observatory observations were made with the NIRC2 instrument on Keck-II behind the natural guide star AO system.  The observations were made on 2017~Aug~20 in the narrow-band $Br-\gamma$ filter in the standard 3-point dither pattern that is used with NIRC2 to avoid the left lower quadrant of the detector which is typically noisier than the other three quadrants. The dither pattern step size was $3\arcsec$ and was repeated three times, with each dither offset from the previous dither by $0.5\arcsec$.  The observations utilized an integration time of 3 seconds with one coadd per frame for a total of 27 seconds.  The camera was in the narrow-angle mode with a full field of view of $10\arcsec$ and a pixel scale of approximately $0.1\arcsec$ per pixel. The Keck AO observations clearly detected a faint companion approximately $0.7\arcsec$ to the south of the primary target.  However, good relative photometry of the detected companion was hampered by the fixed speckle pattern, which our post-processing was unable to fully remove.

EPIC~247589423 was re-observed with the $200\arcsec$ Hale Telescope at Palomar Observatory on 2017~Sep~06 utilizing the near-infrared AO system P3K and the infrared camera PHARO \citep{hayward2001}.  PHARO has a pixel scale of $0.025\arcsec$ per pixel with a full field of view of approximately $25\arcsec$. The data were obtained with a narrow-band $Br$-$\gamma$ filter $(\lambda_o = 2.166; \Delta\lambda = 0.02\mu$m ), a narrow-band $H$-continuum filter $(\lambda_o = 1.668; \Delta\lambda = 0.0018\mu$m ), and a standard $J$-band filter $(\lambda_o = 1.246; \Delta\lambda = 0.162\mu$m). 

The AO data were obtained in a 5-point quincunx dither pattern with each dither position separated by 4$^{\prime\prime}$.  Each dither position is observed 3 times with each pattern offset from the previous pattern by $0.5^{\prime\prime}$ for a total of 15 frames.  The integration time per frame was 4.2 seconds, 9.9 seconds, and 1.4 seconds in the $Br$-$\gamma$, $H$-cont, and $J$ filters. We use the dithered images to remove sky background and dark current, and then align, flat-field, and stack the individual images. The PHARO AO data have a resolution of 0.10$^{\prime\prime}$ (FWHM) in the $Br$-$\gamma$ filter and 0.08$^{\prime\prime}$ (FWHM) in the $H$-$cont$ and $J$ filters, respectively.

The sensitivities of the AO data were determined by injecting fake sources into the final combined images with separations from the primary targets in integer multiples of the central source's FWHM \citep{furlan2017}.  The sensitivity curves shown in Figure \ref{fig:ao_contrast} represent the 5$\sigma$ limits of the imaging data.

The nearby stellar companion was detected in all three filters with PHARO.   The companion separation was measured from the $Br$-$\gamma$ image and found to be $\Delta\alpha = 0.10\arcsec \pm 0.003\arcsec$ and $\Delta\delta = 0.723\arcsec \pm 0.03\arcsec$.  At the distance of the Hyades, the companion has a projected separation from the primary star of $\approx 40$ AU.  The AO imaging rules out the presence of any additional stars within $\sim 0.5$\arcsec\ of the primary ($\sim 30$~AU) and the presence of any brown dwarfs, or widely-separated tertiary components beyond 0.5\arcsec ($\sim 30-1000$~AU).  The presence of the blended companion must be taken into account to obtain the correct transit depth and planetary radius \citep{ciardi2015}.

Table~\ref{tab:stellar} presents the deblended magnitudes of both stars. The stars have blended 2MASS magnitudes of $J = 9.343 \pm 0.026$ mag, $H=8.496 \pm 0.02$ mag and $K_s = 9.196 \pm 0.023$ mag.  The stars have measured magnitude differences of $\Delta J = 4.97 \pm 0.04$ mag, $\Delta H = 4.96 \pm 0.03$ mag, and $\Delta K_s = 4.65 \pm 0.03$ mag. $Br$-$\gamma$ has a central wavelength that is sufficiently close to $Ks$ to enable the deblending of the 2MASS magnitudes into the two components.  The primary star has deblended real apparent magnitudes of $J_1 = 9.11 \pm 0.04$ mag, $H_1 = 8.51 \pm 0.02$ mag, and $Ks_1 = 8.38 \pm 0.02$ mag, corresponding to $(J-H)_1 = 0.060 \pm 0.05$ mag and $(H-K_s)_1 = 0.13 \pm 0.02$ mag; the companion star has deblended real apparent magnitudes of $J_2 = 14.1 \pm 0.1$ mag, $H = 13.47 \pm 0.04$ mag, and $Ks_2 = 13.03 \pm 0.03$ mag, corresponding to $(J-H)_2 = 0.63 \pm 0.11$ mag and $(H-K_s)_2 = 0.44 \pm 0.05$ mag. Utilizing the $(Kepmag - Ks)\ vs.\ (J-Ks)$ color relationships \citep{howell2012}, we derive approximate deblended Kepler magnitudes of the two components of $Kepmag_1 = 10.9\pm0.1$ mag and $Kepmag_2 = 17.4\pm0.2$ mag, for Kepler magnitude difference of $\Delta Kepmag = 6.5\pm0.2$ mag, which is used when fitting the light curves and deriving a true transit depth.

\begin{figure}[!ht]
\begin{center}
\includegraphics[angle=0,scale=0.4,keepaspectratio=true]{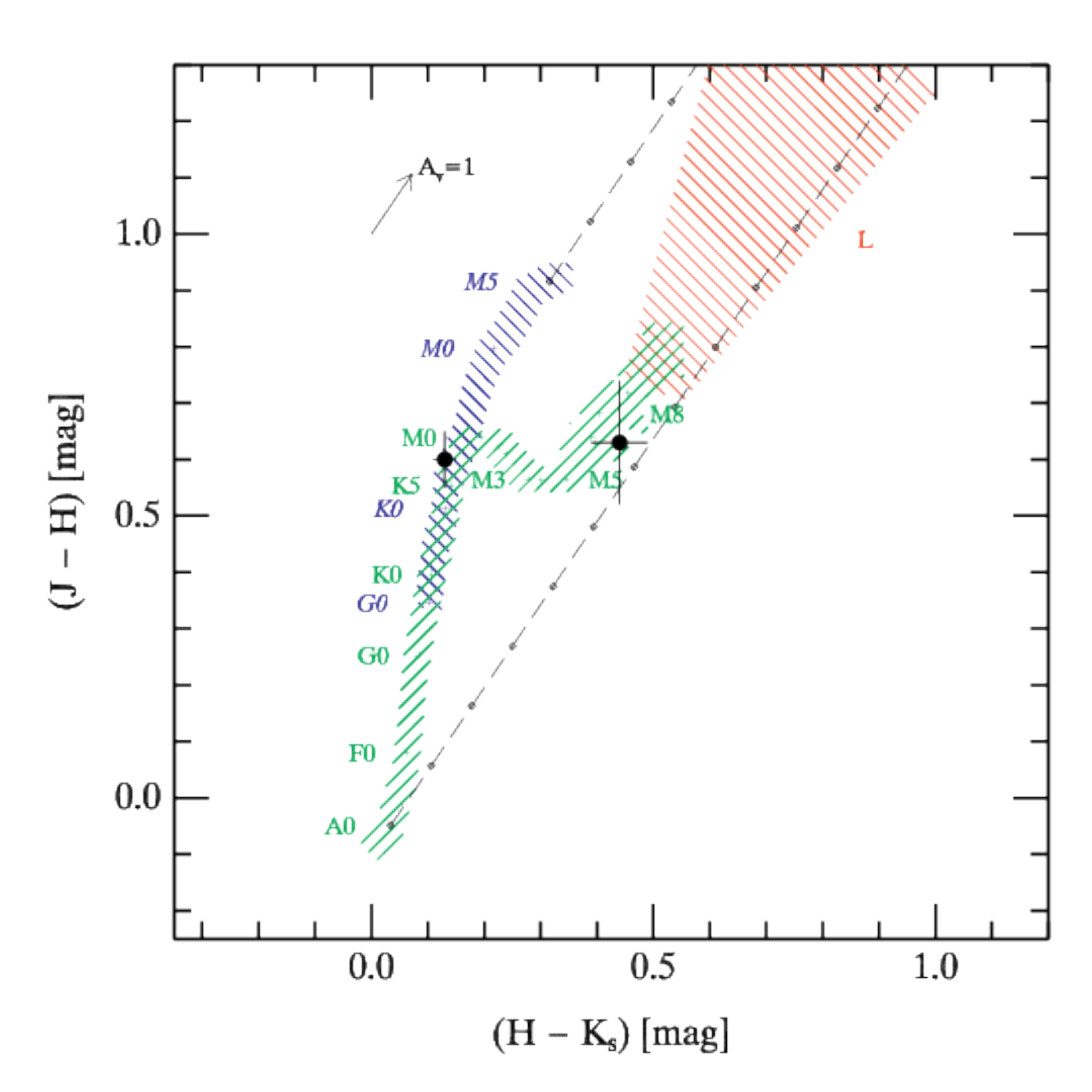}
\caption{2MASS $JHKs$ color-color diagram showing the dwarf branch locus (green), the giant branch locus (blue), and the brown dwarf locus (red).  The black dashed lines represent the direction of reddening induced by extinction ($A_V$).  The positions of the stellar components are overplotted showing the primary component is consistent with being a K5V, and the secondary component is consistent with being an M7/8V.\label{fig:cc}}
\end{center}
\end{figure}

The companion star has infrared colors that are consistent with M7/8V spectral type (Figure~\ref{fig:cc}).  It is unlikely that the star is a heavily reddened background star. Based upon an $R=3.1$ extinction law, an early-F or late-A star would have to be attenuated by more than 6 magnitudes of extinction to make the star appear as a late M-dwarf.  The entire line-of-sight extinction through the Galaxy is only $A_V\approx2$~mag \citep{sf2011} making a background A or F star an unlikely source of the detected companion. 

\section{Association with Hyades Cluster}\label{sec:hyades}
There is a sparse amount of literature on our target star, but it has been consistently regarded as a Hyades member. The star was first proposed as a Hyades member by \citet{weis1983} on the basis of photometry and proper motions, and included in a later study on the H-R diagram of the cluster \citep{reid1993}. The star was also detected as an X-ray source from ROSAT observations \citep{stern1995} and as a GALEX NUV source, possibly due to the low-mass companion which we report here. Finally, it was included in a previous search for transiting planets in the Hyades using photometry from the WASP telescope, and indeed reported as a candidate transiting planet host \citep{gaidos2014}. However, the period and depth of the candidate signal detected by those authors ($P$=3.169 d, $\delta$=0.38\%) bears no resemblance to any transit or stellar variability signal observed in the K2 photometry. 

The current Gaia release (DR1) only has a photometric magnitude ($G=10.4$ mag) for the primary star and has no detection for the companion star. The association of the stars with the Hyades cluster can be investigated via photometric and/or kinematic methods.

The spectroscopic observations (\S\ref{subsec:spec}) and the infrared colors of the primary star are consistent with the primary star being a K5V (see Figure~\ref{fig:cc}).  In the V-band, the M7/8V companion is expected to be more than 10 magnitudes fainter; as a result, the measured optical magnitude of $V=11.20\pm0.03$ mag is dominated by the primary star at the 99.99\% level.  Thus, the V-band magnitude can be utilized to determine the photometric distance to the primary star.  

Based upon the 625--800 Myr isochrone models from \citet{choi2016}, a K5V star has an absolute magnitude of $M_V=7.57$, corresponding to a distance of $d_{phot}\sim 53\pm1$ pc.  Given the $10-20$~pc spread in the Hyades cluster \citep{mann2016}, this distance is in reasonable agreement with the cluster center distance of 45 pc.

The kinematics of the Hyades cluster center have been re-evaluated with the release of the Gaia DR1 and have the following values for the cluster center radial and proper motions: $v_{rad}=39.1\pm0.02$ km/s, $\mu_\alpha = 104.92 \pm 0.12$ mas/yr, and $\mu_\delta = -28.00 \pm 0.09$ mas/yr.  The values measured for EPIC~247589423 very similar to those of the Hyades cluster center (Table~\ref{tab:stellar}). 

Using the measured proper motions from UCAC4 and the radial velocity derived from the HIRES spectrum ($v_{rad} = 39.6\pm0.2$ km/s, $\mu_\alpha = 81.8 \pm 1.0$ mas/yr, $\mu_\delta = -35.2 \pm 0.9$ mas/yr), we recalculated the $UVW$ components for the target, but allowed the distance to vary from 1~pc to 100~pc in steps of 1~pc.  By minimizing the differences between the derived $UVW$ velocities and those established for the Hyades cluster center \citep{vanleeuwen2009}, we derived a kinematic distance of $d_\mathrm{kin} = 58\pm2$~pc. We also used the star's partial kinematics and the methods presented in \citet{lepine2009} to calculate the predicted radial velocity of the star if it is a Hyades member. We find RV$_p$ = 37.8 $\pm$ 0.9 km s$^{-1}$, consistent with our measured HIRES RV at the 2$\sigma$ level. With the general agreements between the photometric and kinematic distances and the general agreement with the kinematic parameters and distance of the Hyades cluster center, we regard EPIC~247589423A as a Hyades cluster member with $>90\%$ probability.

The association of the M7/8V companion to the cluster can only be based upon photometric considerations.  The absolute magnitudes of a late M-dwarf (M7/8) star span $M_J \approx 10-11$ mag and $M_K \approx 9-10$ mag corresponding to a distance for the detected M-dwarf companion of $d \sim 40-60$ pc \citep{choi2016}. The photometrically derived distances are consistent with the average distance to the Hyades and with the distance to primary K5V star.  While not definitive, the spatial coincidence and the similar distances of the K5V and M7/8V stars suggests that the M-dwarf companion may be a physically associated star, and EPIC~247589423 is a wide binary system. Additional high-resolution imaging will be required to demonstrate common proper motion and physical association.

\begin{figure}[!ht]
\begin{center}
\includegraphics[angle=0,scale=0.6,keepaspectratio=true]{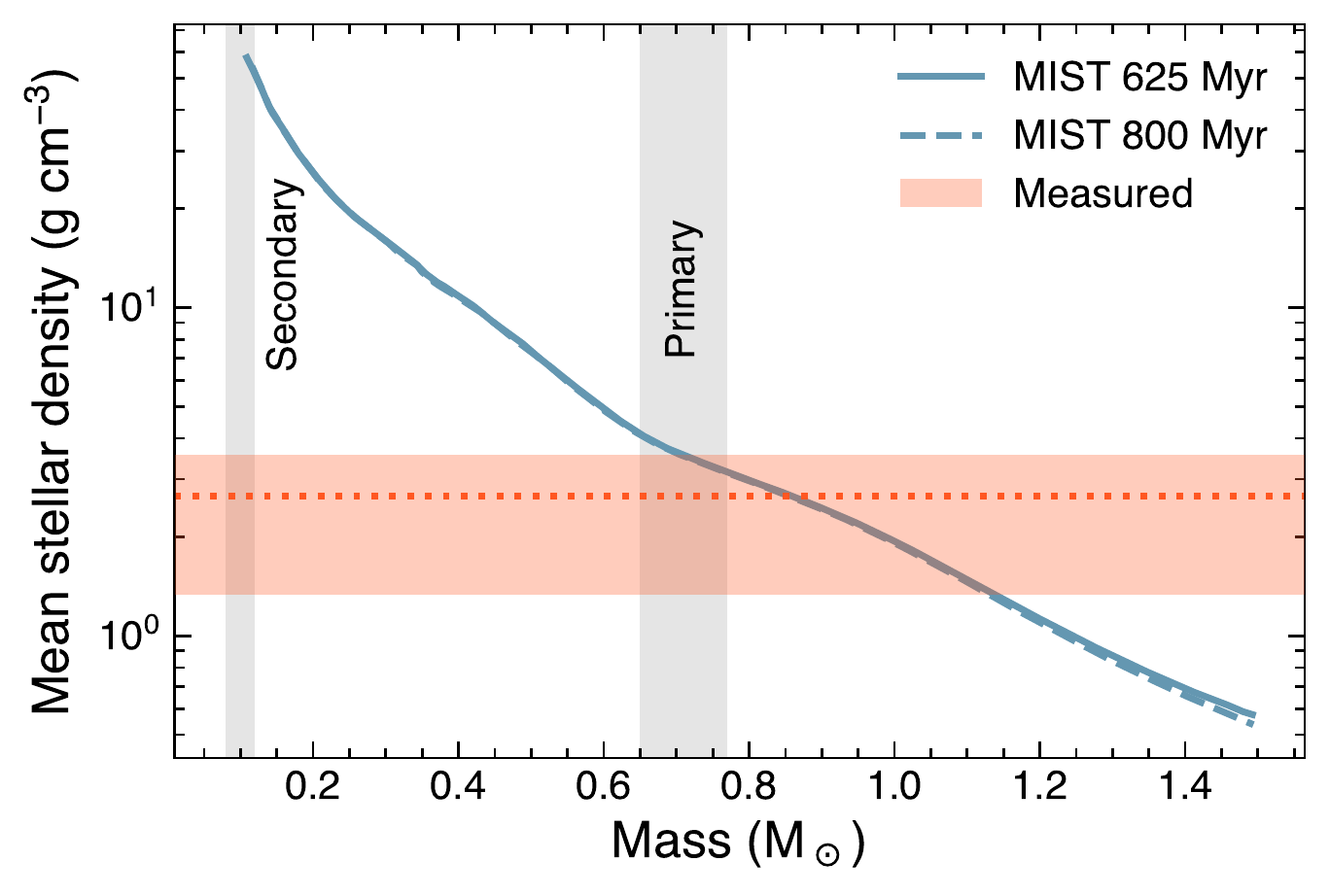}
\caption{Stellar density as a function of the stellar mass as derived from the \citet{choi2016} models for a 625 Myr and an 800 Myr set of isochrones, both with [Fe/H]=+0.13, appropriate for the Hyades.  The horizontal dashed line and associated shaded area represent the derived stellar density and uncertainties from the transit fit (assuming a circular orbit). The vertical gray lines indicate the adopted primary mass and approximate secondary mass. \label{fig:density}}
\end{center}
\end{figure}

\section{Discussion}\label{sec:discussion}
\subsection{Neptune-sized Planet Orbiting the Primary Star}\label{subsec:planet}
The large 4\arcsec\ pixels of Kepler mean we cannot isolate the transit to either the primary or secondary star, from the Kepler data alone. However, we rule out the possibility that the observed transits are of the M7/M8 star due to the lack of a secondary eclipse. We also show that the transit duration strongly favors a planet transiting the K5 star and not the M7/8 dwarf companion. 

With a flux difference in the Kepler bandpass of 6.5 magnitudes, the transit/eclipse would have to be $\gtrsim 65\%$ deep in order to be occurring around the M7/8, after the dilution of the brighter K5V is taken into account.  Such a deep eclipse would require the late-type star to be an eclipsing binary star system, which would require a secondary eclipse depth of  $\lesssim 25\%$.  Even with the dilution of the primary star, a 25\% eclipse would still produce an observed eclipse that is $\approx 1\%$ deep.  Yet, no secondary eclipse (to a limit of $\approx 0.03\%$) is observed, indicating that the transit event is not a stellar eclipse around the companion M-star.  

Additionally, the observed transit duration is more consistent with the orbiting event being around the primary K5V star rather than the M7/8V companion. The time between first and last contact is $T_{14}=3.59\pm0.15$~hr. For a K5V star and the measured stellar radius of $R\sim0.7$~R$_\odot$ and mass of $M \sim0.7$~M$_\odot$, a circular orbit with a period of 17.3 days would have a transit duration of $T_{14}\approx3.7$~hr.  If instead the star that is transited is the M7/8V star, the stellar radius and mass reduce to $R\sim0.1$~R$_\odot$ and $M\sim0.1$~M$_\odot$, and corresponding transit duration would only last 1~hr -- significantly shorter than the observed transit duration \citep{smo2003}.

The transit duration could, of course, be longer if the orbit is not circular:
\begin{equation}
\frac{t_\mathrm{ecc}}{t_\mathrm{circ}} = \frac{\sqrt{(1-e^2)}}{1+e\cos(\omega-90^\circ)}
\end{equation}
where $e$ is the eccentricity and $\omega$ is the argument of periastron \citep[e.g.,][]{kane2012}.  In order to achieve a transit duration near what is observed ($\sim3.5$~hr), the eccentricity would need to be $e\gtrsim0.85$ and the transit would need to occur near apoapsis.  For any other argument of periastron, the eccentricity would need to be even larger.  While this is not impossible, it seems a rather contrived scenario; of the 876 confirmed planets with eccentricity estimates, only 14 have a eccentricities of $e\gtrsim0.8$ and only 7 confirmed planets have an eccentricity of $e\gtrsim0.8$.  Given that none of these systems have orbital periods less than 70 days, and these systems represent only $1-2\%$ of the 876 confirmed planets with measured eccentricities, we consider such a scenario for the planet presented here as unlikely.

Finally, the stellar density from the transit duration of the light curve is more consistent with the host star being a K5V star than being an M7/8V star.  Based upon the \citet{choi2016} models, a mid- to late-M-dwarf with a mass of $M \sim0.1$~M$_\odot$ should have a stellar density near $\rho \gtrsim 30$~g~cm$^{-3}$. By comparison, the stellar density for a K5V star with a mass of $M \sim0.7$~M$_\odot$ should be near $\rho \approx 3.5$~g~cm$^{-3}$, and is in reasonable agreement with the derived stellar density, assuming a circular orbit (see Figure~\ref{fig:density}). We also note that the higher end of the measured stellar density distribution is most consistent with our adopted primary mass and radius.  

We also applied the \texttt{vespa} planet validation tool to this system. This tool assumes that a planet candidate orbits a single main-sequence star, so here we  assume that the planet orbits the brighter of our two stars and that stars in the Hyades have converged onto the main sequence.  In this analysis, which also incorporates our high-resolution imaging data and our exclusion of additional spectroscopic companions, \texttt{vespa} returns a false positive probability of $8\times10^{-5}$. Because of the caveats already mentioned, we do not take this as the true false positive probability, but qualitatively it indicates that, if the planet orbits the brighter K5V star, then it is likely not a false positive.   

We regard all these items -- the lack of a secondary eclipse, the length of the transit duration, the agreement of the derived stellar density with that of a K5V star, and the \texttt{vespa} results -- as sufficient evidence to indicate that the observed transit most likely occurs around the primary star, that it is caused by a planet, and that, given the transit depth and the stellar radius ($0.71$R$_\odot$), the transiting planet is Neptune-sized.

As in our team's previous work \citep{schlieder2016,crossfield2017, dressing2017b} we use the free \texttt{BATMAN}\footnote{\url{https://github.com/lkreidberg/batman}} software \citep{kreidberg2015} to derive transit parameters from our light curve. We ran light curve fits while imposing Gaussian priors on the limb-darkening coefficients, using values appropriate for stars of K5V stars and included the dilution of the transit caused by the blending of the K5V with the M7/8V star. The derived transit and planet parameters are presented in Table~\ref{tab:planet}, and the final fit to the phase-folded light curve is shown in Fig.~\ref{fig:fits}d.  Based upon the transit fits and the HIRES stellar parameters, the transit is caused by a Neptune-sized planet (Rp$=3.03^{+0.53}_{-0.47}$ R$_\oplus$) orbiting the K5V primary star with an orbital period of P$=17.3077\pm0.0013$ days.  

We now refer to the planetary system as the following separate components: K2-136A is the primary K5V star and K2-136B is the M7/8V stellar companion.   In the course of writing this paper, the authors became aware of a similar discovery paper \citep{mann2017b}. In that paper, they report the simultaneous discovery of the Neptune-sized planet reported here. They derive a very similar planetary radius ($R_p\approx2.9\pm0.1\ R_\oplus$ {\em vs.} $R_p\approx3.0\pm0.5\ R_\oplus$). In addition, they report two other planets in the system: an inner earth-sized planet ($R_p=0.99\pm0.05\ R_\oplus$) and an outer superearth-sized planet ($R_p=1.45\pm0.1\ R_\oplus$). In their paper, they ``letter'' the planets in order of orbital period: the inner planet as K2-136~b, the outer planet as K2-136~d, and the Neptune-sized planet, jointly discovered, is referred as K2-136~c.  We adopt the same lettering scheme in this paper; however, given our discovery of the stellar companion, the planets should should be referred to K2-136A~b, K2-136A~c, and K2-136A~d.

\subsection{Stellar Rotation Period and Alignment}\label{subsec:rotation}
The light curve is clearly modulated by stellar variability that appears to be quasi-periodic (Fig.~\ref{fig:fits}b).  The full amplitude of the variations is $\sim0.5\%$ ($\sim 5$~mmag) which is comparable to field K-dwarfs \citep{ciardi2011}.  Being $\sim 6.5$ magnitudes fainter than the K-dwarf, the M-dwarf companion would need to have variability amplitudes on the order of $1-2$ magnitudes in order to produced the observed amplitude of variability. That level of  variability associated with quasi-periodic rotation is typically not observed in the field or Hyades M-dwarfs \citep{ciardi2011,douglas2016}. Thus, the (primary) source of the observed variability is likely the primary component of the system: K2-136A.

A Lomb-Scargle periodogram of the light curve shows its strongest peak at $15.2\pm0.2$~d, and an autocorrelation of the light curve shows its strongest (non-zero-lag) peak at $13.8\pm1.0$~d. Such a period is consistent with the periods of other Hyades members of a similar mass \citep{delorme2011,douglas2016} and further evidence that the spot modulation pattern is due to the primary, rather than the secondary which would be expected to be rotating more rapidly; it therefore seems possible that the stellar rotation period of K2-136A lies in this range. 

A rotation period of $14-15$ days is expected to produce an equatorial velocity for an $0.71$ R$_\odot$ of $V_\mathrm{eq}\approx 2.4-2.5$ km/s. The HIRES spectrum yields \vsini\,=\,$3.9\pm1.0$~km~s$^{-1}$, which is marginally consistent with the expected equatorial velocity derived from the rotation periods. The  modest inconsistencies between our measured \vsini\ and expectations from the stellar radius and photometric rotation period might be accounted for by (1) systematic effects involved in our estimation of \vsini, of order 1~km~s$^{-1}$, and (2) surface differential rotation. Measured and expected differential rotation rates in K-dwarfs are in the range of $\lesssim$0.05~rad~d$^{-1}$ \citep{barnes2005, kitchatinov2012}. If the modulation pattern in the K2 photometry is due to surface features at higher, more slowly rotating, latitudes, it is possible the equatorial rotation period is shorter by $\sim$1~d. 

While not definitive, the marginal agreement between the measured \vsini\ and the rotation period indicates that the star's rotational axis is nearly perpendicular to the orbital plane of K2-136A~c.  If there were a significant misalignment, we would expect a more significant difference between the light curve derived rotation period and the measured \vsini, although a longer time baseline would be useful to confirm this. 

Finally, we note that the orbital period of the planet and the rotation period of the star are similar, but not the same.   We estimated how long the planet would take to circularize ($\tau_\mathrm{circ}$) using the equation given by \citet{al2006}:

\begin{equation}
\begin{split}
\tau_\mathrm{circ} = 1.6\ Gyr \times \left (\frac{Q_p}{10^6}\right )
    \times \left (\frac{M_*}{M_\odot}\right )^{-1.5}\\
    \times \left (\frac{M_p}{M_\mathrm{Jup}}\right )
    \times \left (\frac{R_p}{R_\mathrm{Jup}}\right )^{-5}
    \times \left (\frac{a}{0.05 AU}\right )^{6.5}
\end{split}
\end{equation}

The tidal circularization time scales linearly with $Q_p$, and the tidal parameter ($Q_p$) is notoriously uncertain. However, the Neptune value is estimated to be $\sim 10^5$ with a possible range of $10^4 - 10^6$ \citep{maness2007}, indicating that the circularization timescale for K2-136A~c may be $\sim500 - 600$ Myr -- a timescale very similar to the age of the Hyades Cluster.  If the tidal parameter is more akin to Jupiter ($10^6$), the circularization timescale would be closer to 5~Gyr, well beyond the age of the Hyades.    

\begin{figure}[!ht]
\begin{center}
\includegraphics[angle=0,scale=0.3,keepaspectratio=true]{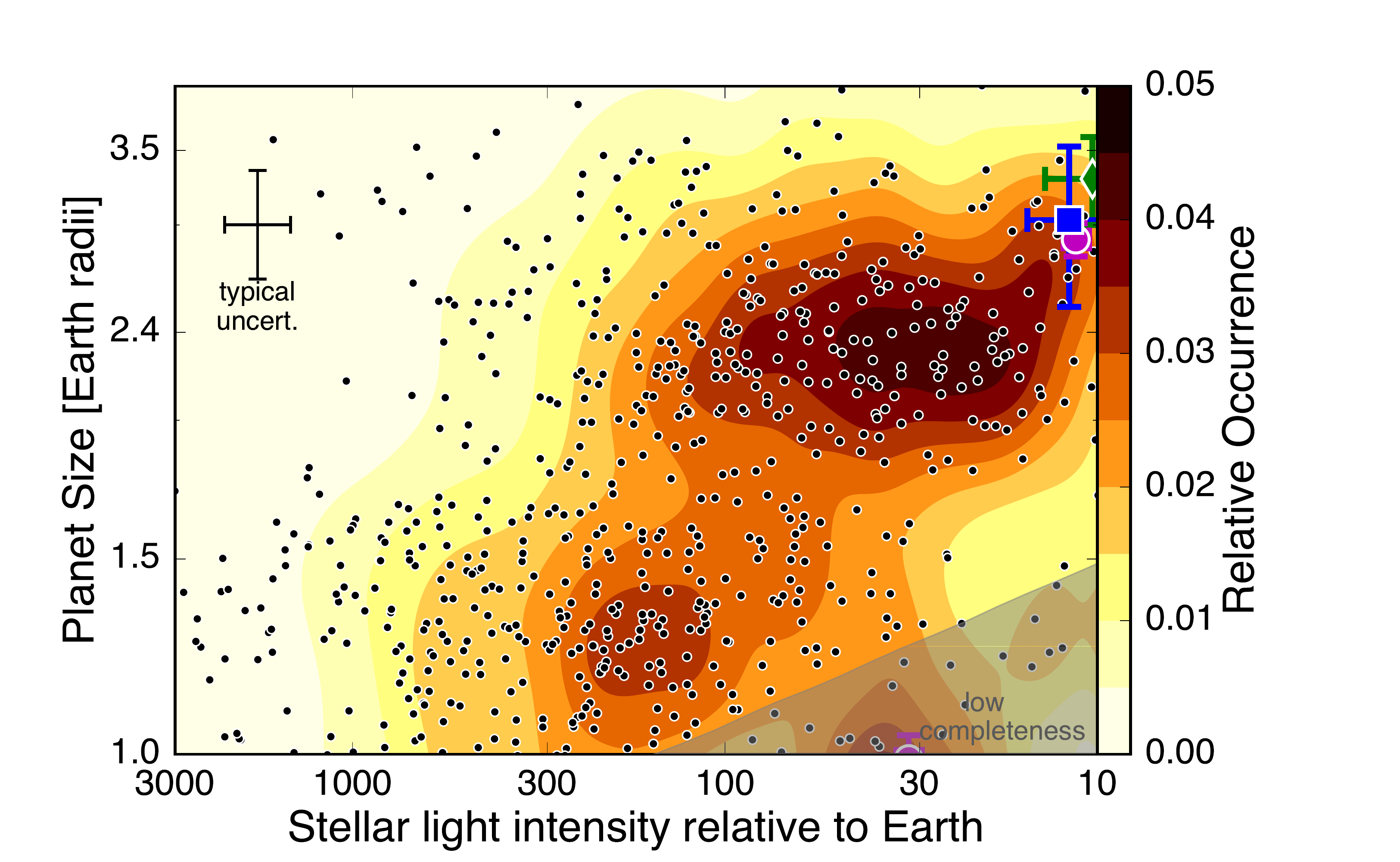}
\caption{The two-dimensional distribution of planet size and incident stellar flux, adopted from \citet{fulton2017}, is shown with the location of the previously known transiting planet in the Hyades Cluster: K2-25b (green diamond) and the planets in K2-136.  The blue square represents K2-136A~c from this work and the magenta circles represent the K2-136 planets as measured by \citet{mann2017b}.  K2-136A~d is off the right-side of the figure with an insolation flux of $\approx 5$. \label{fig:evaporation}}
\end{center}
\end{figure}

\begin{figure}[!ht]
\begin{center}
\includegraphics[angle=0,scale=0.6,keepaspectratio=true]{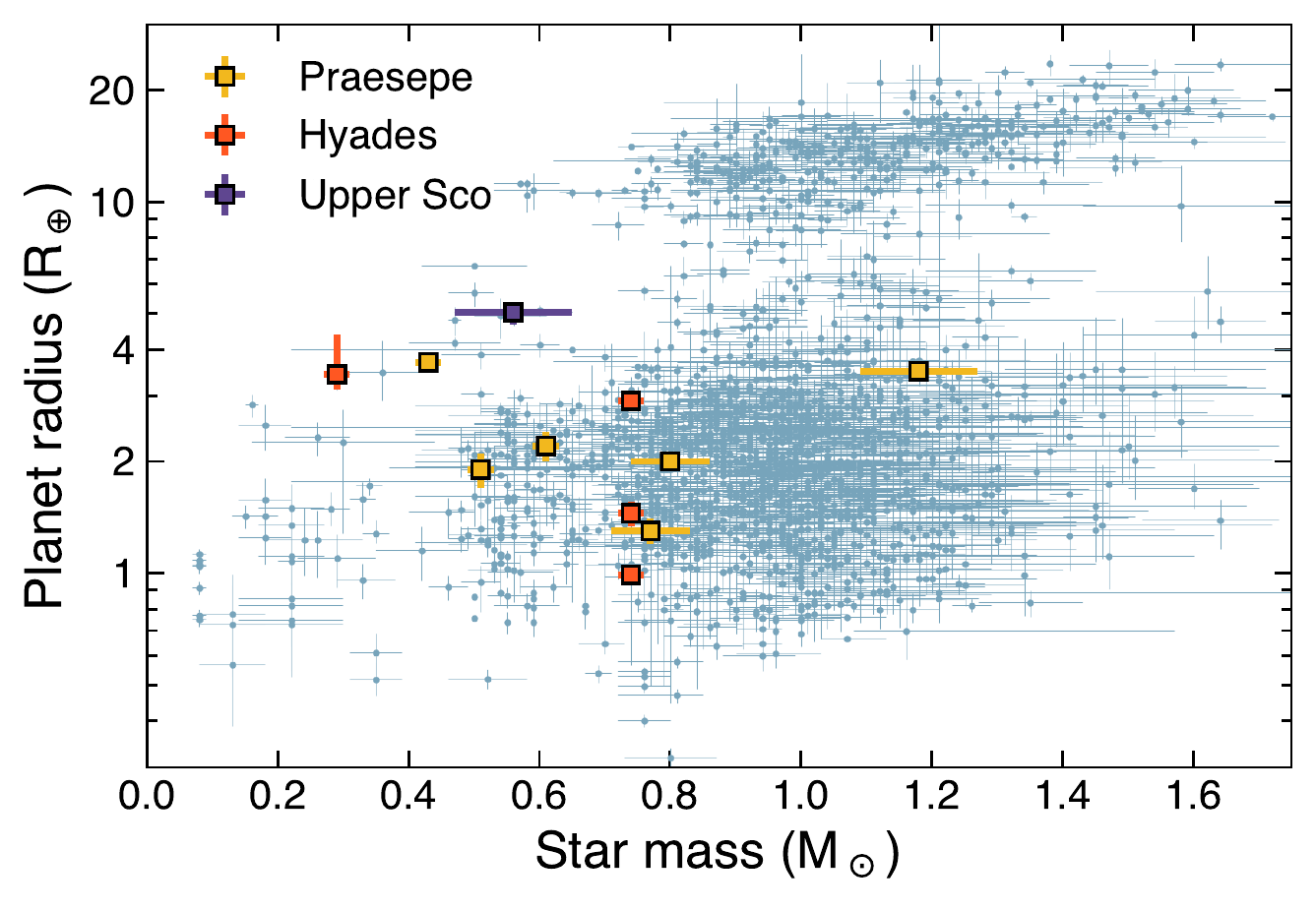}
\caption{The two-dimensional distribution of planet size and stellar mass.  The positions of the known cluster transiting planets from the Praesepe, Upper Sco, and Hyades clusters are shown.  K2-136A~c is marked with the red star. The blue points represent the known transiting planets with orbital periods of $\le 30$ days. Data were gathered from the NASA Exoplanet Archive.\label{fig:rad_vs_mass}}
\end{center}
\end{figure}
\subsection{Comparison to Other Systems}\label{subsec:compare}
K2-136A~c is in a 17 day orbit around a K5V star and is experiencing a stellar insolation flux of S$\sim 12$ S$_\oplus$ (Table~\ref{tab:planet}) and as a result it has an expected equilibrium temperature that is $T_\mathrm{eq} = 400 - 600$~K depending on the planet albedo and the atmosphere re-circulation.  The other previous detected transiting planet in the Hyades \citep[K2-25b,][]{mann2016} orbits an M4.5V star, but has a much shorter orbital period of 3.485 days but experiences a similar insolation flux (S$\sim 10$ S$_\oplus$) as K2-136A~c.

The two Neptunes-sized planets in K2-25 and K2-136, respectively, occupy a similar location in the two dimensional distribution of planet radius \textit{vs.} stellar insolation flux (see Figure~\ref{fig:evaporation}).  Both planets are at the edge of the distribution, perhaps indicating that the $\sim0.6-0.8$ Gyr Hyades planets, in comparison to the $>$Gyr Kepler sample used to define the evaporation valley, may still need to undergo significant evolution. 

If the known young cluster transiting planets are compared to the old field star transiting planets (primarily dominated by Kepler detections) in a two dimensional distribution of planet radius \textit{vs.} stellar mass (see Figure~\ref{fig:rad_vs_mass}), the distribution of the cluster planets does not look significantly different than the distribution of old field planets.  This suggests that the long term evolution of the cluster planets should lead them to the distribution of planets currently observed in the field.   

\subsection{Potential for More Follow-Up}\label{subsec:followup}
K2-136A~c orbits a relatively bright star in the infrared ($K\sim 9$ mag) in a very well-studied and nearby open cluster and, thus, offers the opportunity for more detailed studies (e.g., with Spitzer and JWST). In the optical, the star is a bit  fainter with $V \approx 11$ magnitude.  If the K2-136A~c has a similar density to Neptune, the expected radial velocity (RV) amplitude caused by the orbital motion of the planet should be on the order of 5~m~s$^{-1}$, well within the reach of modern radial velocity spectrographs. As noted above, the system has been detected as an X-ray source which may indicate that the star is active; however, this could be the M-dwarf companion and not the K-dwarf primary.  The spectra of the K-dwarf appears to have an activity indicator of $S_{HK}=1.03$, suggesting that the RV jitter could be of 1--10~m~s$^{-1}$ \citep{isaacson:2010}. Thus an RV measurement of the planet's mass may be feasible.  Such a measurement would provide an all-too-rare constraint on the bulk properties of a young sub-Neptune. 

Unfortunately, the ecliptic latitude of K2-136A is $\sim 1^\circ$ and (like most  K2 targets) it will not be observed in the prime TESS mission.  However, the transit depth is approximately 1.5 mmag and, thus, ground-based observations of the transits may be possible to refine the transit ephemeris and to search for long-term timing variations indicative of the other planets in the system \citep[e.g.,][]{lendl:2017,barros:2017}.

\section{Summary}\label{sec:summary}
We present the discovery of a sub-Neptune-sized ($3.0$ R$_\oplus$) planet in a 17.3 day orbit around a K-dwarf in the Hyades cluster.  The host star also appears to have a late M-dwarf companion that is separated from the primary star by at least 40 AU.  This planetary system, K2-136A~c, represents the fourth planet discovered in the Hyades cluster, and only the second transiting planet in the Hyades.  Both transiting planets now known in the Hyades are Neptune-sized and orbit relatively low-mass stars; K2-25b orbits an M4.5V dwarf and the newly presented K2-136A~b orbits a K5V dwarf which also has two other smaller planets and low-mass M-dwarf companion.  

By finding and studying planets in clusters spanning a range of stellar ages, we may begin to understand how and on what timescales planetary systems form and evolve. The planets discovered in the Upper Sco, Praesepe, and Hyades clusters provide snapshots in time and represent the first steps in mapping out this evolution. As we begin to understand the planetary distribution in the nascent clusters in which stars and their planetary systems are born, we can begin to set constraints on and understand how planetary systems form and evolve into the systems we see today in the field of stars.

\acknowledgments
The authors thank the Andrew Mann and his collaborators for contacting us regarding their efforts so that we could work together to submit our respective discovery papers.  We also note that another paper was submitted after the submission of our paper which is consistent with the results presented here \citep{livingston2017}.  The authors thank the referee for comments which help to improve the clarity of the manuscript.

The authors wish to recognize and acknowledge the very significant cultural role and reverence that the summit of Maunakea has always had within the indigenous Hawaiian community. We are most fortunate to have the opportunity to conduct observations from this mountain. This research has made use of the NASA Exoplanet Archive and the ExoFOP website, which are operated by the California Institute of Technology, under contract with the National Aeronautics and Space Administration under the Exoplanet Exploration Program. MB acknowledges support from the North Carolina Space Grant Consortium. LA acknowledges support from NASA's Minority University Research and Education Program Institutional Research Opportunity to the University of the Virgin Islands. BT acknowledges support from the National Science Foundation Graduate Research Fellowship under grant number DGE1322106 and NASA's Minority University Research and Education Program. Finally, DRC would like to dedicate this paper to Teresa Ciardi for her years of insight to all of my papers - and this paper was no exception.

%\facilities{Kepler, IRTF, Keck:II, Hale}

\newpage
\begin{deluxetable}{l r r }[bt]
\hspace{-1in}\tabletypesize{\scriptsize}
\tablecaption{  Stellar Parameters \label{tab:stellar}}
\tablewidth{0pt}
\tablehead{
\colhead{Parameter} & \colhead{Value} & \colhead{Notes}
}
\startdata
\multicolumn{3}{c}{\em Identifying Information} \\
EPIC ID & 247589423 & \\ %& \cite{huber2016} \\
$\alpha$ R.A. (hh:mm:ss) & 04:29:39.0 & Gaia\\
$\delta$ Dec. (dd:mm:ss) & +22:52:57.8 & Gaia\\
$\mu_{\alpha}$ (mas~yr$^{-1}$) & $+81.8\pm 1.0$ & UCAC4\\
$\mu_{\delta}$ (mas~yr$^{-1}$) & $-35.2\pm0.9$ & UCAC4\\
Barycentric RV (km~s$^{-1}$)  & $39.6\pm0.2$ & HIRES; This Work\\
$S_{HK}$  &         1.027 & HIRES; This Work\\
Distance (pc) & $50-60$ & This Work\\
Age (Myr) & $625 - 750$ &  \citet{perryman1998}\\ 
          &             &  \citet{bh2015}\\
\multicolumn{3}{c}{\em Blended Photometric Properties} \\
NUV (mag) ........  & $19.47  \pm 0.10$  & GALEX \\
B (mag) ..........  & $12.479 \pm 0.041$ & APASS \\
V (mag) ..........  & $11.200 \pm 0.030$ & APASS \\
g (mag) ..........  & $11.969 \pm 0.030$ & APASS \\
r (mag) ..........  & $10.746 \pm 0.040$ & APASS \\
Kepmag (mag)        & $10.771$           & \citet{huber2016} \\
i (mag) ..........  & $10.257\pm 0.020$ & APASS \\
J (mag) ..........  & $9.096 \pm 0.022$ & 2MASS \\
H (mag) ..........  & $8.496 \pm 0.020$ & 2MASS \\
Ks(mag) .........   & $8.368 \pm 0.019$ & 2MASS \\
\\
\multicolumn{3}{c}{\em Deblended Photometric Properties} \\
\underline{K5V Star}\\
Kepmag (mag)        & $10.9 \pm 0.1$ & A-Component \\
J (mag) ..........  & $9.11 \pm 0.04$ & A-Component \\
H (mag) ..........  & $8.51 \pm 0.02$ & A-Component \\
Ks(mag) .........   & $8.38 \pm 0.02$ & A-Component \\
\underline{M7/8V Star}\\
Kepmag (mag)        & $17.4\pm 0.2$ & B-Component \\
J (mag) ..........  & $14.1 \pm 0.1$ & B-Component \\
H (mag) ..........  & $13.47 \pm 0.04$ & B-Component \\
Ks(mag) .........   & $13.03 \pm 0.03$ & B-Component \\
\\
\multicolumn{3}{c}{\em A-Component Spectroscopic Properties}\\
Spectral Type & K5V $\pm$ 1 & SpeX\\
\Teff\ (K) & $4364 \pm 70$ & HIRES\\
           & $4360 \pm 206$& SpeX\\
$[$Fe/H$]$ & +0.15 $\pm$ 0.09 & HIRES\\
$M_*$ ($M_\odot$) & $0.71 \pm 0.06$ & HIRES\\
                  & $0.70 \pm 0.07$  & SpeX \\
$R_*$ ($R_\odot$) & $0.71 \pm 0.10$ & HIRES\\
                  & $0.67 \pm 0.06$ & SpeX\\
$L_*$ ($L_\odot$) & $0.164 \pm 0.031$   & HIRES\\
                  & $0.152 \pm 0.052$   & SpeX\\
$\log_\mathrm{10} g$ (cgs) & $4.63 \pm 0.11$ & HIRES\\
                           & $4.62^{+0.05}_{-0.10}$ & SpeX\\
\vsini\ (km~s$^{-1}$) & $3.9\pm 1.0$  &  HIRES
\enddata
\tablenotetext{}{HIRES stellar parameters used for transit modeling}
\end{deluxetable}

\begin{deluxetable}{llll}[bt]
\tabletypesize{\scriptsize}
\tablecaption{  Planet Parameters \label{tab:planet}}
\tablewidth{0pt}
\tablehead{
\colhead{Parameter} & \colhead{Symbol} & \colhead{Units} & \colhead{Value} 
}
\startdata
Time of Transit Center &   $T_{0}- 2454833$     & $BJD_\mathrm{TDB}  $ & $2997.0235\pm0.0025$ \\
Orbital Period & $P$ &  d & $17.3077\pm0.0013$\\
Orbital Inclination & $i$ & deg & $89.30^{+0.49}_{-0.76}$\\
Planet/Star Radius Ratio & $R_P/R_*$ & \% & $3.85^{+0.47}_{-0.20}$\\
Linear Limb Darkening & $\alpha$ & -- & $0.900\pm0.030$\\
Quadratic Limb Darkening & $\beta$ & -- & $0.486\pm0.030$\\
Transit Duration ($1^{st}-4^{th}$) & $T_{14}$ & hr & $3.59^{+0.17}_{-0.14}$\\
Transit Duration ($2^{nd}-3^{rd}$) &  $T_{23}$ & hr & $3.22^{+0.15}_{-0.18}$\\
Stellar Radius-Orbit Ratio &  $R_*/a$ & -- & $0.0287^{+0.0075}_{-0.0027}$\\
Impact Parameter &  $b$ & -- & $0.43\pm0.28$\\
Stellar Density & $\rho_{*,circ}$ & g~cm$^{-3}$ & $2.67^{+0.90}_{-1.34}$\\
Semi-major Axis &   $a$ & AU & $0.11728\pm0.00048$\\
Planet Radius & $R_P$ & $R_\oplus$ & $3.03^{+0.53}_{-0.47}$\\
Incident Flux & $S_{inc}$ & $S_\oplus$ & $11.9^{+3.7}_{-3.2}$\\
Secondary Eclipse Depth & $\delta_\mathrm{ecl}\ (3\sigma)$ &        ppm & $< 238$

\enddata
\end{deluxetable}

\end{document}